\documentclass[11pt]{article}
\usepackage{amsmath}
\usepackage{graphicx,psfrag,epsf}
\usepackage{enumerate}
\usepackage{natbib}
\usepackage{url} 
\usepackage{bm}
\usepackage{amsthm}% not crucial - just used below for the URL 
\usepackage{setspace}
\usepackage{color}
\usepackage{slashbox}
\usepackage{verbatim}
\usepackage{times} 

\newcommand{\argmax}{\textrm{argmax}}
\newcommand{\Cor}{\textrm{Cor}}

\renewcommand{\vec}[1]{\boldsymbol{#1}}

\newcommand{\bigCI}{\mathrel{\text{\scalebox{1.07}{$\perp\mkern-10mu\perp$}}}}

\newtheorem{proposition}{Proposition}
\newtheorem{corollary}{Corollary}
\newtheorem{definition}{Definition}

\newcommand{\E}{\textrm{E}}

\newcommand{\blind}{0}

\begin{document}

\def\spacingset#1{\renewcommand{\baselinestretch}%
{#1}\small\normalsize} \spacingset{1}

%%%%%%%%%%%%%%%%%%%%%%%%%%%%%%%%%%%%%%%%%%%%%%%%%%%%%%%%%%%%%%%%%%%%%%%%%%%%%%

\if0\blind
{
  \title{\bf Emulation of utility functions over a set of permutations: sequencing reliability growth tasks}
  \author{Kevin J Wilson, Daniel A Henderson\\
    School of Mathematics and Statistics, Newcastle University, UK\\
    and \\
    John Quigley \\
    Department of Management Science, University of Strathclyde, UK}
  \maketitle
} \fi

\if1\blind
{
  \bigskip
  \bigskip
  \bigskip
  \begin{center}
    {\LARGE\bf Emulation of utility functions over a set of permutations: sequencing reliability growth tasks}
\end{center}
  \medskip
} \fi

\bigskip
\begin{abstract}
We consider Bayesian design of experiments problems in which we maximise the prior expectation of a utility function over
a set of permutations, for example when sequencing a number of tasks
to perform. When the number of tasks is large and the expected utility
is expensive to compute, it may be unreasonable or infeasible to
evaluate the expected utility of all permutations. We propose an
approach to emulate the expected utility using a surrogate function
based on a parametric probabilistic model for permutations. The surrogate function is fitted by maximising the correlation with the expected utility over a set of training points. We propose a suitable transformation of the expected utility to improve the fit. We provide results linking the correlation between the two functions and the number of expected utility evaluations to undertake. The approach is applied to the sequencing of reliability growth tasks in the development of hardware systems, in which there is a large number of potential tasks to perform and engineers are interested in meeting a reliability target subject to minimising costs and time. An illustrative example shows how the approach can be used and a simulation study demonstrates the performance of the approach more generally.  Supplementary materials for this article are available online.
\end{abstract}

\noindent%
\vfill
\hfill {\tiny technometrics tex template (do not remove)}

\newpage
\spacingset{1.45} % DON'T change the spacing!

\section{Introduction}

The maximisation of the prior expectation of a utility function to
find the optimal design of an experiment in a Bayesian analysis is
challenging as a result of the necessity
of evaluating the expected utility for all designs. We can
make progress by using an approximation based on a surrogate function which is faster to evaluate, known as an emulator \citep{Ken01,Hen09,You11,Zho11}. When the possible experimental designs are a set of permutations this adds complexity as commonly used emulators, such as Gaussian process emulators, are based on the assumption of a smooth relationship between the inputs to, and outputs from, the function being emulated, which is unlikely to be true in this case.

An example is the reliability growth of hardware products under development. During development, initial designs are subject to detailed analysis, identifying improvements until performance requirements are met. Reliability tasks which analyse the design and facilitate the enhancement include fault tree analysis, failure modes and effects analysis and highly accelerated life testing. These tasks can be resource intensive and costly to implement and each has the goal of design improvement by understanding weaknesses. Multiple tasks may uncover the same fault in the current design of the product and some faults may not be uncovered by any of the tasks. Therefore, the engineers are faced with the decision problem of which tasks to perform and in what order to grow their reliability to the required level. Their possible experimental designs are the permutations of the possible reliability tasks to carry out.

\subsection{Emulation of permutations}

The use of surrogate models for emulating expensive functions dates
back at least as far as \cite{SacksWMW89} who propose a Gaussian
process prior for the output of a complex computer model. An early
example of emulating an expensive utility function in the context of
Bayesian optimal design is described in \cite{MuellerP95}. The use of
surrogate models for emulating an expensive function over a set of
permutations has received relatively little attention. An
early example is described in \cite{VoutchkovKBO05} in which a
surrogate model is proposed for ordering a sequence of welding tasks
in the aircraft industry. Most surrogate models for functions with
continuous input spaces, such as Gaussian processes or radial basis
functions, are based on the premise that points close in input space
will lead to similar values of the expensive function.  In terms of a
generic approach to the problem, a natural first step for emulating a
function over permutations would be to replace the Euclidean distance
between inputs in surrogate models with a more natural distance
measure on the space of permutations. This is the approach proposed by
\cite{MoraglioK11,MoraglioKY11} and \cite{KimMKY14}. These authors
focused on a radial basis function-based surrogate model for general
combinatorial input spaces, including permutations. They used the
Hamming distance and Kendall tau distance as two distance measures on
sets of permutations; see \cite{Mar95} for details.

This approach of replacing the Euclidean distance with a
permutation-based distance was generalised by
\cite{ZaeffererSFFNB-B14} to a Gaussian process-based surrogate model.
The use of a Gaussian process-based surrogate allowed the authors to
use the efficient global optimisation approach of \cite{JonesSW98} and a standard Kriging estimator. The authors reported improved
performance over a radial basis function-based surrogate on a set of
combinatorial optimisation problems.  This work was extended by
\cite{ZaeffererSB-B14} who investigated 14 different distance
functions over sets of permutations for Gaussian process-based
surrogates. An approach based on choosing the distance function by
maximum likelihood yielded the best results, with the Hamming distance
the best single distance measure. \cite{Zaefferer15} describes an R
package `CEGO' for implementing the Gaussian process-based surrogate
modelling and optimisation approach that is proposed in
\cite{ZaeffererSFFNB-B14} and \cite{ZaeffererSB-B14}.

\subsection{Design for reliability}

Design for reliability principles are not sufficient for designing complex systems \citep{Way15,Dir11,Def08} and as such reliability growth continues to provide a key role in product development.  However, achieving growth through Test Analyse And Fix (TAAF) is expensive which leads organisations to develop integrated programs of activities to reduce reliance on testing \citep{Kra04}.  There are several activities employed to enhance reliability during design and development, e.g. Highly Accelerated Life Testing (HALT), Fault Tree Analysis, Failure Modes and Effects Analysis (FMEA) (see \cite{Bli11}), each of which requires resources and must be managed, see for example IEC 61014.  While these activities have different perspectives upon a design, each has the common goal of seeking to improve the design.  To date, little attention has been paid in the literature on optimally sequencing activities to achieve target reliability while minimising costs.  

Past research has focused primarily on either determining optimal designs, for example with levels of redundancy \citep{Lev15,Cas15}, or modelling the fault detection rate from TAAF tests, for which there is international standard IEC 61164, comprising both frequentist  and Bayesian methods \citep{Cro74,Wal99,Qui03}.  Research at the program level is scarce, either focusing on monitoring progress graphically \citep{Kra15,Wal05} or project management issues such a metrics, infrastructure and documentation \citep{Eig15}.  While minimising program costs has been explored by \cite{Hsi03a,Hsi03b} a key shortcoming has been the representation of growth and program costs through continuous functions as decisions are discrete and choices are finite \citep{Gui02}.  An exception to this is \cite{Joh06} who explored minimising program cost with discrete choices through integer programming methods and \cite{Wil16} who provided a solution incorporating multi-attribute utility functions to allow trade-offs between attributes; however neither paper addressed the issue of sequencing the activities.  

Table \ref{rgex} provides a simplified illustrative example which is similar to practice but substantially reduced in size and presented without industrially sensitive design details.  Table \ref{rgex} presents five engineering concerns with the design of an aerospace system and four development activities; these were identified through an elicitation process.

\begin{table}[ht]
\centering
\begin{tabular}{|c|cccc|} \hline
\backslashbox{Concern}{Activity} &	Salt and Fog &	Vibration Test & Electrical &	Highly Accelerated  \\ 
& Stress Test & & Stress Test & Life Test (HALT) \\ \hline
Temperature control &&&& \\ 				
Moisture Ingress	&&&& \\			
Difference in TCE &&&& \\				
Circuitry voltage spike &&&& \\				
Impedance &&&& \\ \hline 
\end{tabular}
\caption{Illustrative efficacy matrix with 5 concerns and 5 activities }
\label{rgex}
\end{table}

The ability of the design to maintain its temperature within acceptable limits was a concern for the engineers, and while the current design may achieve desired control, the engineers were uncertain.  Moisture ingress was a concern as new material was being used in the design compared with heritage designs and, while preliminary tests indicated that the new material would not be deficient in this regard, the engineers were uncertain.  The significance of the difference of the Temperature Coefficient of Expansion (TCE) between different materials was a concern for aspects of the system as this may prove to be a deficiency in the design. The possibility of aspects of the circuitry generating unacceptable voltage spikes was identified.  The final concern listed related to uncertainty between the soldering, the environment and the circuits, where residue remaining on solder joint could cause failure for high impedance circuits in environments of high moisture.  Associated with each concern is a probability that the concern is a fault, i.e. if the system were used in operation it would fail due to an imperfection or deficiency described by the concern.

There are a variety of reliability tests that can be conducted to assess an item's design at the component, sub-assembly or system level; these include life and environment tests, for example Highly Accelerated Life Testing (HALT), as well as environmental testing against specific stresses such as vibration, heat, electricity, salt, fog or moisture. For a more detailed discussion see \cite{Oco12} or \cite{Sil98}.  Table \ref{rgex} presents four such activities and within each cell of the table we would input a probability measuring how likely the associated activity would be to identify the associated concern as a fault, assuming that the concern was a fault, i.e. the probability it does not slip through test undetected. 

Once populated with probabilities, efficacy matrices can help identify if there are concerns with no associated activities and can form the basis of a reliability growth programme.  Typically, the matrices have a many to many relationship, where a fault can be exposed by several activities and an activity can expose several faults.  As such, there are choices to be made with respect to which activities should be scheduled, with the aim of either identifying the faults in the design or providing evidence that the concern is not a fault. While we have presented a simple example, a large system design may have hundreds of concerns with as many activities.

\subsection{This paper}

In this paper we initially consider the problem of emulating the prior
expectation of a utility function over a set of permutations. We propose an approach to perform the emulation using the
Benter model \citep{Benter94} as the basis for the surrogate
function. The Benter model is an extension of the Plackett-Luce model
\citep{Plackett75,Luce59}, a popular probabilistic model for
permutations; see \citep{Mar95} for further background on 
probabilistic models for permutations. We outline how to fit the
Benter model based on a training sample of model runs for the expected utility and propose suitable transformations of both the expected utility and surrogate function to improve the fit. We give results which provide us with a method of deciding on the number of evaluations of the expected utility to make based on the optimal values of the surrogate function.

We consider the problem of sequencing reliability growth tasks specifically. We adapt a commonly used model for reliability growth \citep{Joh06,Qui06} under development to incorporate explicitly uncertainty on the reliability function and utilise this to assess the cost and time of performing any sequence of reliability tasks, taking into account that we will stop testing when we reach a reliability target. We propose a two-attribute utility function to solve the design of experiments problem.

%The general form of the reliability growth problem considered is as follows. The goal of the development process is to reach some reliability target, $R_0$. The engineers inside the organisation have a number, $I$, of concerns about the current design of the product, each of which could be a fault which would lead to a failure. There are also a number, $J$, of tasks which could be undertaken to ascertain whether each of the concerns is a fault. Each of the tasks has an associated cost, $Y$, and time, $\chi$. The reliability of the product at time $t$ will depend on whether each of the engineering concerns is a fault and the failure time distributions of each of the faults. The optimal design will be the sequence of tasks which causes the product to meet the reliability target while keeping costs and time low.

In Section \ref{emul} we outline our general approach to emulation of
the expected utilities on sets of permutations and give theoretical
results. In Section \ref{rel} we consider the model for reliability
growth and develop the solution to the decision problem. In Section
\ref{example} we give an illustrative example informed by work with
industrial partners in the aerospace industry. Section \ref{simul}
considers two simulation studies to examine the strengths and weaknesses of the emulator more generally and provide guidance on the choice of training and evaluation set sizes for the emulator. In Section \ref{conc} we summarise and give some areas for further work.

\section{Emulation of utility functions on permutations}
\label{emul}

\subsection{The emulator}
\label{emuldesc}

Suppose that there are $J$ tasks which are to be performed in a
sequence. Then there are $J!$ possible sequences, or permutations, of
the $J$ tasks. Let $\mathcal{S}^J$ denote the set of all permutations
of the $J$ tasks. Each sequence of tasks $\bm
x=(x_1,x_2,\ldots,x_J)\in\mathcal{S}^J$, where $x_k$ denotes the
$k$'th task, gives rise to an expected utility $u=U(\bm x)\in[0,1]$,
where 0 represents the least preferable possible outcome and 1
represents the most preferable possible outcome. 

If we can compute $U(\bm x)$ for all possible sequences then we can solve the system design problem by choosing the sequence which maximises the expected utility. However, it is often the case that $U(\bm x)$ is time consuming to compute so that evaluating the expected utility for each possible permutation $\bm x$ may not be feasible. Instead, we can treat $U(\cdot)$ as an expensive deterministic function and try to ``emulate'' it using a less expensive surrogate, with the idea being that it is feasible to evaluate the surrogate function at all possible permutations in reasonable time, or allow us to explore the space of permutations more efficiently.

Thus we wish to find a surrogate function $f(\cdot)$ which takes a sequence $\bm x$ as input and outputs a real scalar quantity such that
$f(\bm x_i)>f(\bm x_j)~\textrm{if}~U(\bm x_i)>U(\bm x_j)$,
for $i\neq j$. %Ideally, the sequence which maximises the expected utility should also maximise the surrogate function,
%\begin{displaymath}
%\bm\hat{\bm x}=\argmax_{\bm x}U(\bm x)\Rightarrow\bm\hat{\bm x}=\argmax_{\bm x}f(\bm x).
%\end{displaymath}

Our proposed surrogate function,
\begin{equation}
f(\vec{x})\equiv f(\vec{x};\vec{\theta},\vec{\alpha})= \sum_{j=1}^J \alpha_j\log(\theta_{x_j})-\log\left(\sum_{m=j}^J \theta_{x_m}^{\alpha_j}\right),
\label{eq:surr}
\end{equation}
is derived from the loglikelihood function of the Benter model~\citep{Benter94} which we treat as a function of $\vec{x}$ for fixed
values of the positive parameters
$\vec{\theta}=(\theta_1,\ldots,\theta_J)$ and
$\vec{\alpha}=(\alpha_1,\ldots,\alpha_J)$. Note that when $j=J$ the
contribution to $f(\vec{x})$ is 0, and so we cannot estimate
$\alpha_J$; we therefore set $\alpha_J=0$. The Benter model is an
example of a multistage ranking model~\citep{FlignerV88} and was
proposed by Benter to overcome perceived deficiencies of the
Plackett-Luce (PL) model~\citep{Plackett75,Luce59} for analysing the
results of horse races. Benter 
%observed more variability in the lower
%placings, perhaps due to jockeys expending less effort once they
%realised their horse was not going to win any money, and so 
introduced
a parameter $\alpha_j$ for each ``stage'', to
reflect its importance.  The 
stage-dependent flexibility of the Benter model has been used in
several applications, such as the analysis of voting data
\citep{GormleyM08jasa}. A more detailed description of the Benter model, together with other
extensions of the PL model, such as its reversed version, the reverse
Plackett-Luce (RPL) model, is given in \cite{MollicaT14}.  The PL model
is obtained from the Benter model by setting
$\alpha_j=1$ for $j=1,2,\ldots,J$. 
%and we are grateful to one of the
%referees for suggesting the Benter model as an alternative to the PL
%and RPL models which we proposed in a previous version of this paper. 
We are using the parametric structure of the Benter
model simply as a surrogate $f(\vec{x})$ for $U(\vec{x})$ and we are
not implying or assuming that the expected utility is a probability.
As well as displaying good empirical performance, which we detail in Section~\ref{simul}, the parametric structure of the Benter model does have some properties
that may be desirable for emulating the expected utility function of a
set of sequences of tasks. For example, for fixed values of
$\vec{\alpha}$, the positive parameter $\theta_k$ is associated with
task $k\in\{1,2,\ldots,J\}$, such that $\theta_k$ is proportional to
the utility when task $k$ is scheduled first. Similarly, for fixed
$\vec{\theta}$, the positive parameter $\alpha_\ell$ is associated
with position $\ell$ in the sequence of tasks, and can be thought of
as reflecting the relative importance of the task that is performed
$\ell$th in the sequence. We might expect the $\alpha$ parameters to
be larger for the tasks that are performed early on in the sequence
than those performed later in the sequence, when the target
reliability may already have been reached.

We would like to choose
parameters $\vec{\psi}=\{\bm\theta,\vec{\alpha}\}$ such that sequences $\bm x$
with high expected utility $U(\vec{x})$ also have high values of
$f(\vec{x})$. To do so, we choose $\vec{\psi}$ to maximise
the correlation between the logit transformed 
expected utilities from a training sample, $\bm x_1,\ldots,\bm x_N$,
for some $N<<J!$, and their values under the surrogate
$f(\vec{x})$. Note that any appropriate correlation function may be used, e.g.\
Pearson, Spearman, Kendall. Similarly we place no
restrictions on the choice of training sample, but defer discussion of choices of
training sample to Section~\ref{conc}.  Specifically, let
the vector of expected utilities of the training sample be $\bm
u=(u_1,\ldots,u_N)$, where $u_i=U(\bm x_i)$, for
$i=1,2,\ldots,N$; as we observe
expected utilities close to zero and one, we  work with the logit
transformed values of the expected utility,
$\eta_i=\log(u_i/(1-u_i))$. Also, let the vector of surrogate function values
of the training sample be
$\vec{f}^{\bm\psi}=(f_{1}^{\bm\psi},\ldots,f^{\bm\psi}_N)$, where
$f_i^{\bm\psi}=f(\vec{x}_i; \vec{\psi})$ for $i=1,\ldots,N$ with
$f(\cdot;\cdot)$ as defined in Equation~\eqref{eq:surr}. Thus, we seek
$\vec{\psi}$ to maximise $c(\bm\psi)=\Cor(\bm \eta,\bm f^{\bm\psi})$, i.e.,
\begin{displaymath}
\hat{\bm\psi}=\argmax_{\bm\psi}\ c(\bm\psi),
\end{displaymath}
where $\vec{\eta}=(\eta_1,\ldots,\eta_N)$.
The maximisation of the correlation can be performed using the
Nelder-Mead simplex algorithm~\citep{NelderM65} as implemented in the \verb optim  function in \verb R     \citep{R14}. We run the
 optimisation from a small number of different starting points (usually 5), and
 choose the value of $\bm\psi$ which gives the maximum
 correlation.

Our initial emulator of the logit transformed expected utility function
is therefore our surrogate function evaluated at $\hat{\bm\psi}$, that is
$\hat{f}(\cdot)=f(\cdot ;\hat{\bm\psi})$.

\subsubsection{Regression-adjusted surrogate model}
\label{sec:rasurr}

The emulator $\hat{f}(\cdot)=f(\cdot ;\hat{\bm\psi})$ performs well 
empirically, as detailed in Section~\ref{simul}, but it can be improved by
an adjustment based on simple linear regression. We fit a
linear model with response vector $\vec{\eta}$ and 
linear predictor $\beta_0 + \beta_1 \hat{f}(\vec{x}) + \beta_2
\hat{f}(\vec{x})^2 + \beta_3 \hat{f}(\vec{x})^3$  using ordinary least squares, where $\hat{f}(\vec{x})$ is the fitted
value of the surrogate function under the Benter model. 
The resulting
fitted mean function
\[
f^\star(\vec{x}) = \hat{\beta}_0 + \hat{\beta}_1 \hat{f}(\vec{x}) + \hat{\beta}_2 \hat{f}(\vec{x})^2 + \hat{\beta}_3 \hat{f}(\vec{x})^3
\]
 is our regression-adjusted emulator. This
regression-adjusted surrogate improves upon $\hat{f}(\cdot)$ in several
ways. Firstly, it is easier to interpret the output from $f^\star(\cdot)$
as it is on the same scale as the logit-transformed expected
utilities (although we emphasise that this is not an essential feature
of an emulator as all we are interested in is the relative order of
the expected utilities). Secondly, the relationship between logit expected utility
and the surrogate function is made more linear, which aids
interpretation. Thirdly we may obtain a quantification
of the uncertainty in the emulator output for a
given sequence $\vec{x}$ through the usual linear model-based
prediction intervals, if desired. Such prediction
  intervals are necessarily conditional on the estimated surrogate
  model parameters $\hat{\vec{\psi}}$; taking into account the
uncertainty in $\hat{\vec{\psi}}$ would lead to wider intervals. Note that if an emulator for the expected
utility $U(\cdot)$ (rather than the log-transformed expected utility) is
required then we simply take
\[
f^\dagger(\vec{x})=\frac{\exp\{f^\star(\vec{x})\}}{1+\exp\{f^\star(\vec{x})\}},
\]
though such a transformation is not essential. 

%\subsection{Choice of training sample}
%\label{sec:train}

\subsection{Properties of the emulator}

Unless there is a perfect correspondence between $U(\bm x)$ and $f(\bm x)$, there is no guarantee that the optimal sequence under the surrogate function, $\hat{\bm x}$, will be the sequence which maximises the expected utility. In this case, we can use $f(\cdot)$ to propose a set of $M$ candidate sequences which may have a high expected utility, specifically those which maximise $f(\cdot)$. We can then take the sequence out of the $N+M$ evaluations with the largest expected utility as our best estimate of the optimal sequence.

The following results allow us to link the correspondence between $U(\bm x)$ and $f(\bm x)$ and the probability of observing the sequence with maximum expected utility in the $M$ candidate sequences. The proof is given in the Supplementary Material.

\begin{proposition}
\label{prop}
If we have $J$ tasks with $R=J!$ permutations, then the sequence of tasks with highest expected utility will be in the $M$ sequences with highest value of the function $f(\cdot)$ with probability
\begin{displaymath}
\sum_{m=1}^{M}\dfrac{N_{R-1,\delta-m+1}}{N_{R,\delta}},
\end{displaymath}
where $\delta$ is the Kendall's tau distance between the utility and $f(\cdot)$ for all possible sequences, $N_{R,\delta}$ is given by
$N_{R,\delta}=C_{R,\delta}-C_{R,\delta-1}$,
and $C_{R,\delta}$ satisfies the recursion $C_{R,\delta}=\sum_{l=\delta-R+1}^{\delta}C_{R-1,l}$ with $C_{0,0}=C_{i,0}=1$ for $i=1,\ldots,R$.
\end{proposition}
We can relate this to Kendall's correlation. The Kendall correlation between the utility and $f(\cdot)$ for all possible sequences is given by
%\begin{displaymath}
$\tau=(T-2\delta)/T$,
%\end{displaymath}
%for $k=1,\ldots,R$. 
This allows us to consider the probability of observing the sequence with highest expected utility in the $M$ optimal sequences from $f(\cdot)$ directly from the Kendall correlation using the result above.

If the Kendall correlation is strong between the expected utility and the surrogate function, then we can simplify the result further. In particular, if $\delta\leq R-1$, then $N_{R,\delta}=C_{R-1,\delta}$. This leads to the following corollary.

\begin{corollary}
\label{result2}
If $\delta\leq R-1$ then the sequence of tasks with highest expected utility will be in the $M$ sequences with highest value of the surrogate function with probability
\begin{displaymath}
\sum_{m=1}^{M}\dfrac{C_{R-2,\delta-m+1}}{C_{R-1,\delta}}.
\end{displaymath}
\end{corollary}

This will be the case if $\tau\geq 1-\dfrac{2}{T}(R-1)$. We can use the result from the corollary to provide us with a value for $M$ which guarantees that we find the optimal sequence, provided we know $\delta$. The proof is given in the Supplementary Material.
\begin{proposition}
\label{result3}
If $\delta\leq R-1$, then if we choose $M=\delta+1$ the optimal sequence will be in the $M$ sequences with the highest value of $f(\cdot)$ with probability 1. If $M=\delta$, this probability is
\begin{displaymath}
1-\dfrac{1}{C_{R-1,\delta}}.
\end{displaymath}
\end{proposition} 

When the number of items to sequence is large, the value of $M$ needed to guarantee the optimal sequence will be large. Although we can estimate the value of $\delta$ from the training set of sequences, we do not know its population value and so, while these results can give us a feel for a suitable value of $M$, they are not sufficient to allow us to choose $M$. In Section \ref{simul} we provide guidance on choosing the values of $N$ and $M$ based on an extensive simulation study.   

We can illustrate the results using a simple example, which is given in the Supplementary Material.

\section{Sequencing reliability growth tasks}
\label{rel}

\subsection{Reliability of hardware products under development}

We consider the model developed in \cite{Joh06,Qui06}, which is adapted from IEC 61014 and explicates the relationship between the reliability of a system under development and planned development activities. The model is predicated on the concept of concerns, which are possible faults within a system, which may then lead to system failure. The following provides definitions to four key concepts in the model.
\begin{definition}
Reliability is the ability of a system to perform a required function under stated conditions for a stated period of time.
\end{definition}  
\begin{definition}
A failure is the inability of a system to perform a required function under stated conditions for a stated period of time.
\end{definition}  
\begin{definition}
A fault is an imperfection or deficiency in a system such that the system will fail, i.e. not perform a required function under stated conditions for a stated period of time.
\end{definition} 
\begin{definition}
A concern is a possible fault, i.e. imperfection or deficiency in a system such that the system will fail.
\end{definition} 
The reliability of the system is assessed by a probability distribution, which is developed by identifying a set of concerns, assessing the probability that each concern is a fault and specifying distributions describing the time each fault will be realised as a failure. The parameters can be assessed using expert judgement elicitation or historical data \citep{Qui99,Wal01}.

Suppose the concerns associated with the current system design are labelled $i=1,\ldots,I$. Let us define $Z_i$ to be an indicator variable such that $Z_i=1$ if concern $i$ is a fault and 0 otherwise. Define $R_i(t)$ to be the probability that concern $i$ would result in failure after time $t$ conditional on concern $i$ being a fault. If the realisations of failures are independent then the reliability of the system can be expressed as
$$R(t,\bm z)=\prod_{i=1}^{I}[1-z_i(1-R_i(t))],$$
where $\bm Z$ is the vector of indicator variables for all concerns. For each concern we have an associated probability of the concern being a fault, elicited from expert judgement elicitation, $\lambda_i=\Pr(Z_i=1)$, so the expectation of $R(t,\bm z)$ with respect to $\bm Z$ is easily obtained by substituting $z_i$ for $\lambda_i$, under the assumption that $Z_i\bigCI Z_{i^{'}}$, for $i\neq i^{'}$.

During a reliability development program activities are performed
assessing the design of the system, where the outcome of each activity
is either to confirm a concern as a fault or provide evidence to the
contrary. We assume that once a fault has been confirmed it is designed
out of the system. The reliability growth during the program is
captured through Bayesian updating of $\lambda_i$ given test
data. We define the following two indicator variables. Denote $\kappa_j$ to represent whether activity $j$, for
$j=1,\ldots,J$, has been conducted or not, i.e. $\kappa_j=1$ or 0
respectively.  Let $D_{i,j}$ indicate whether concern $i$ is realised as a fault in activity $j$ ($D_{i,j}=1$), or not ($D_{i,j}=0$).

We define $p_{i,j}$ as the conditional probability associated with $D_{i,j}$, $p_{i,j}=\Pr(D_{i,j}=1\mid Z_i=1)$, and this is is specified through expert elicitation based on the idea of an efficacy matrix \citep{Joh06,Wil16}.

It is sufficient for the model to utilise the indicator $D_i=\max_{j=1,\ldots,J}(\kappa_j D_{i,j})$. Assuming that test outcomes are independent, the updated probability of a concern being a fault given a test program has not realised the concern as a fault, through Bayes Theorem, is
\begin{displaymath}
\Pr(Z_i=z_i\mid D_i=0)=\begin{cases}
\dfrac{1-\lambda_i}{1-\lambda_i[1-\prod_{j=1}^{J}(1-p_{i,j})^{\kappa_j}]},~z_i=0, \\
\dfrac{\lambda_i\prod_{j=1}^{J}(1-p_{i,j})^{\kappa_j}}{1-\lambda_i[1-\prod_{j=1}^{J}(1-p_{i,j})^{\kappa_j}]},~z_i=1, \\
\end{cases}
\end{displaymath}
with $\Pr(Z_i=1\mid D_i=1)=1$ and $\Pr(Z_i=0\mid D_i=1)=0$ as, when an item fails due to concern $i$, it must be a fault.

This allows us to evaluate the prior expectation of the reliability, given a planned set of development activities. The reliability in this case can be expressed as $R(t,\bm z)=\prod_{i=1}^{I}R_i(t)^{I[z_i>d_i]}$, where $I[z_i>d_i]$ is an indicator function which takes the value 1 if $z_i>d_i$ and 0 otherwise, under the assumption that once a fault has been identified in a task it is designed out of the system with probability one.

\begin{eqnarray} \nonumber
\E_{\bm D}\left\{\E_{\bm Z\mid \bm D}\left[R(t,\bm z)\right]\right\} %& = & \prod_{i=1}^{I}\left[\sum_{d_i=\{0,1\}}\Pr(D_i=d_i)\right. \\ \label{rel}
%&& \left.\times\sum_{x_i=\{0,1\}}\Pr(X_i=x_i\mid D_i=d_i)R_i(t)^{I[x_i>d_i]}\right] \\
& = & \prod_{i=1}^{I}\left[1-\left(1-R_i(t)\right)\lambda_i\prod_{j=1}^{J}(1-p_{i,j})^{\kappa_{j}}\right],
\end{eqnarray}
where $\bm D$ is the vector of indicator variables $D_i$. A full derivation of this result is given in the Supplementary Material.

\subsection{Optimal sequencing of reliability growth tasks}

We could perform all of the possible reliability growth tasks for a system. Following this, if the system meets some pre-specified reliability target $R_0$, it will be released. However, if we were to reach our reliability target after fewer than the allocated tasks, then we would stop testing and save time and money. Therefore, finding the optimal sequence of reliability tasks is an important question.

We need to consider the probability distribution of the reliability, as we are interested in quantities of the form $\Pr(R(t,\bm z)\geq R_0)=\alpha(t)$. We transform the reliability so that it is not restricted to $[0,1]$. Specifically,
\begin{displaymath}
g(t,\bm z)=\log\left[R(t,\bm z)\right].
\end{displaymath}
Assume that the transformed reliability follows a Normal distribution $g(t,\bm z)\sim N(m(t),v(t))$. Justification for this choice will be given later. Then the probabilities of interest are $\Pr(R(t,\bm z)\geq R_0)=\Pr(g(t,\bm z)\geq\log R_0)$.

We can fully specify the distribution of the reliability by specifying $m(t),v(t)$ as $m(t)=\E_{\bm D}\left\{\E_{\bm Z\mid\bm D}\left[g(t,\bm z)\right]\right\}$ and $v(t)=\E_{\bm D}\left\{\E_{\bm Z\mid\bm D}\left[g(t,\bm z)^2\right]\right\}-\E_{\bm D}\left\{\E_{\bm Z\mid \bm D}\left[g(t,\bm z)\right]\right\}^2$. %These are calculated via
%\begin{displaymath}
%\E_{\bm D}\left\{\E_{\bm Z\mid\bm D}\left[g(t,\bm z)^n\right]\right\}=\left[\sum_{\bm d}\prod_{i=1}^{I}\Pr(D_i=d_i)\sum_{\bm z}\prod_{i=1}^{I}\Pr(Z_i=z_i\mid\bm d)\right]\times\left[\sum_{i=1}^{I}\log\left(R_i(t)\right)\right]^n,
%\end{displaymath}
%for $n=1,2$.

These are not fast and efficient calculations to perform. Each vector $\bm d$ and $\bm z$ are of length $I$ with each element having two possible states, 0 and 1. The number of sequences of length $J$ is $J!$. Therefore the total number of calculations which would be necessary to evaluate the probability distribution of $g(t,\bm z)$ for all sequences of length $J$ is $G=2^{2I+1}\times J!$. For example, if $I=5,J=5$ then $G=245,760$ and if $I=15,J=14$ (still reasonably small) then $G=1.87\times10^{20}$. In practice, for a reasonably large problem, it is not going to be possible to evaluate all of the required expectations exactly.

We can approximate the logarithm of the reliability by taking a rare event approximation to obtain
\begin{displaymath}
-\log\left[R(t,\bm z)\right]=\sum_{i=1}^{I}(1-R_i(t))z_i-\left[\sum_{i=1}^{I}(1-R_i(t))z_i\sum_{j=1}^{J}\kappa_j(1-\mu_{i,j})\right],
\end{displaymath} 
where $\mu_{i,j}$ is 1 if task $j$ finds fault $i$ given that fault $i$ exists and 0 if it does not find $i$ when it exists. In this case,
\begin{eqnarray*}
m(t) & = & -\left[\sum_{i=1}^{I}(1-R_i(t))\lambda_i-\left(\sum_{i=1}^{I}(1-R_i(t))\lambda_i\sum_{j=1}^{J}\kappa_j(1-p_{i,j})\right)\right], \\
v(t) & = & \sum_{i=1}^{I}\left[(1-R_i(t))\lambda_i\right]^2\sum_{j=1}^{J}\kappa_j(1-p_{i,j})p_{i,j}.
\end{eqnarray*}
By the Lyapanov Central Limit Theorem \citep{Kni00} the distribution of the approximation is asymptotically Normal with increasing numbers of activities.

Using the approximations the total number of calculations required to solve the design problem for a sequence of length $J$ reduces to $H=2IJ\times J!$. In the specific cases given above, if $I=5,J=5$ then $H=6000$ and if $I=15, J=14$ then $H=3.66\times 10^{13}$. We see that the number of calculations required has been significantly reduced.

\subsection{Bayesian expected utility solution}

We define a utility function which incorporates the uncertainty on the reliability $R(t,\bm z)$. The multi-attribute utility function we define will include this uncertainty in both the conditional utilities for financial cost and time cost. Thus the multi-attribute utility function will depend on attributes $(Y,\chi)$, where $Y$ represents financial cost and $\chi$ represents time.

Recall that the possible sequences of tasks of length $J$ are given by $\bm x=(x_1,\ldots,x_J)$ and that $\bm x\in\mathcal{S}^J$, where $\mathcal{S}^J$ represents the set of all of the permutations of sequences of length $J$. Then the Bayesian optimal sequence of tasks is given by
\begin{displaymath}
\argmax_{\bm x\in\mathcal{S}^J}\left[\E_{\bm D}\left\{\E_{\bm Z\mid\bm D}\left[U(\bm x,\bm z)\right]\right\}\right],
\end{displaymath} 
where the utility function $U(\bm x,\bm z)$ incorporates the probability distribution of $R(t,\bm z)$ in the following way.

Suppose that, for sequence $\bm x$, the costs associated with the individual tasks are $(y(x_1),\ldots,y(x_J))$ and the times are $(\chi(x_1),\ldots,\chi(x_J))$. Then if we perform $j$ tasks the total costs and times will be given by $y_j^{(tot)}(\bm x)=\sum_{k=1}^j y(x_k)$ and $\chi_j^{(tot)}(\bm x)=\sum_{k=1}^j \chi(x_k)$ respectively. Introducing the rule that we stop testing if we reach the target reliability, the cost and time for sequence $\bm x$ are
\begin{eqnarray*}
C(\bm x,\bm z)  =  \sum_{j=1}^{J}\left[y_j^{(tot)}(\bm x)\gamma_j\prod_{k=1}^{j-1}(1-\gamma_j)\right],~
T(\bm x,\bm z)  =  \sum_{j=1}^{J}\left[\chi_j^{(tot)}(\bm x)\gamma_j\prod_{k=1}^{j-1}(1-\gamma_j)\right],
\end{eqnarray*}
%We can define a linear utility solution to the optimal sequencing problem as we did for the optimal allocation problem. In this case we can define the marginal utility of reliability to be
%\begin{displaymath}
%U(R)=\sum_{j=1}^{J}\gamma_j,
%\end{displaymath}
where $\gamma_j$ is an indicator variable which takes the value 1 if $R(t,\bm z)>R_0$ and 0 if not. If we assume utility independence between cost and time we can then define the general utility function to be a binary node:
\begin{displaymath}
U(\bm x,\bm z)=q_1U(C(\bm x,\bm z))+q_2U(T(\bm x,\bm z))+q_3U(C(\bm x,\bm z))U(T(\bm x,\bm z)),
\end{displaymath}
for trade-off parameters $q_1\geq0, q_2\geq0$ and $-q_i\leq q_3\leq 1-q_i$ for $i=1,2$ such that $q_1+q_2+q_3=1$. Examples of suitable risk averse marginal utility functions for financial cost and time would be 
\begin{equation}
U(C(\bm x,\bm z))=1-(C(\bm x,\bm z)/Y_0)^2,~~U(T(\bm x,\bm z))=1-(T(\bm x,\bm z)/\chi_0)^2,
\label{eq:util}
\end{equation}
where $Y_0$ is the maximum budget and $\chi_0$ is the maximum time to carry out the tasks, however any suitable functions on $[0,1]$ could be used. %The financial and time costs of testing are given by $Y=\sum_{j=1}^{J}Y_j(1-\gamma_{j-1})$ and $\chi=\sum_{j=1}^{J}\chi_j(1-\gamma_{j-1})$, where $\gamma_0=0$. 
%Linear utility for financial and time costs are then
%\begin{displaymath}
%U(Y)-1-\dfrac{Y}{Y_0}, ~ U(\chi)=1-\dfrac{\chi}{\chi_0}.
%\end{displaymath}

We see that, to calculate the expected utility of a particular sequence, we need $\E[\gamma_j]$, for all $j$, which are the $\Pr(R(t,\bm z)>R_0)$ for a particular stage of a specific sequence.

Although the rare event approximation reduces the number of calculations required to evaluate the probability distribution of the reliability, and hence the expected utilities of the sequences of tasks, even for a small problem with $I=15, J=14$ the number of calculations required to solve the decision problem is not feasible to undertake in practice. Instead, we propose to use the surrogate functions developed in Section \ref{emul} to approximate the expected utilities and solve the decision problem approximately. This approach is illustrated using an example and evaluated using a simulation study in the next two sections.

%\subsection{Emulation of the expected utility}

%In the notation of Section \ref{emul}

\section{Illustrative example}

\label{example}

\subsection{Background}

Suppose that in an elicitation engineers identify 15 concerns in a product under development. There are 9 possible tasks which the engineers could carry out to identify if these concerns are faults and, if they are, design them out. This means that in all there are 362,880 possible sequences of the tasks which could be carried out.

Each task has associated with it a cost of between 0 and 50 units and a duration of between 0 and 20 units. The target reliability, which would be assessed by the decision maker, is 0.8, the maximum time is 150 units and the maximum total cost is 132 units. If the conditional rate of failure for concern $i$, given that it is a fault, can be thought of as constant over time a suitable choice for the time to failure is the exponential distribution.  This results in a reliability function of  $R_i(t)=\exp\left\{-\epsilon_i t\right\},$
where $\epsilon_i$ is the rate of failures resulting from concern $i$, given that it is a fault. The final parameters which need to be specified are the trade-off parameters for the binary utility function. In the example, the $\lambda_i$ are between 0 and 0.5, approximately 50\% of the $p_{i,j}$ are equal to zero indicating that task $j$ will not find concern $i$, given that it is a fault, and the rest are between 0 and 0.5 and each $\epsilon_i$ is chosen to be 0.02. The values of $\lambda_i$ and $p_{i,j}$ used in the example are given in Table \ref{priors}.

\begin{table}[ht]
\centering
\begin{tabular}{|c|c|ccccccccc|} \hline
$i$ & $\lambda_i$ & $p_{i,1}$ & $p_{i,2}$ & $p_{i,3}$ & $p_{i,4}$ & $p_{i,5}$ & $p_{i,6}$ & $p_{i,7}$ & $p_{i,8}$ & $p_{i,9}$ \\ \hline
1 & 0.13 & 0.00 & 0.19 & 0.47 & 0.00 & 0.00 & 0.00 & 0.00 & 0.00 & 0.25 \\
2 & 0.19 & 0.00 & 0.40 & 0.00 & 0.00 & 0.39 & 0.26 & 0.01 & 0.00 & 0.24 \\
3 & 0.29 & 0.22 & 0.00 & 0.00 & 0.33 & 0.00 & 0.00 & 0.00 & 0.00 & 0.04 \\
4 & 0.45 & 0.17 & 0.17 & 0.00 & 0.43 & 0.00 & 0.22 & 0.20 & 0.00 & 0.00 \\
5 & 0.10 & 0.00 & 0.00 & 0.00 & 0.39 & 0.23 & 0.00 & 0.33 & 0.00 & 0.32 \\
6 & 0.45 & 0.00 & 0.00 & 0.49 & 0.18 & 0.00 & 0.00 & 0.00 & 0.50 & 0.00 \\
7 & 0.47 & 0.00 & 0.00 & 0.00 & 0.29 & 0.00 & 0.46 & 0.28 & 0.49 & 0.34 \\
8 & 0.33 & 0.12 & 0.00 & 0.00 & 0.05 & 0.31 & 0.16 & 0.00 & 0.00 & 0.00 \\
9 & 0.31 & 0.00 & 0.22 & 0.44 & 0.00 & 0.00 & 0.00 & 0.43 & 0.00 & 0.32 \\
10 & 0.03 & 0.30 & 0.15 & 0.00 & 0.44 & 0.00 & 0.47 & 0.36 & 0.00 & 0.14 \\
11 & 0.10 & 0.06 & 0.16 & 0.13 & 0.00 & 0.00 & 0.28 & 0.00 & 0.48 & 0.00 \\
12 & 0.09 & 0.41 & 0.00 & 0.48 & 0.00 & 0.00 & 0.46 & 0.13 & 0.00 & 0.17 \\
13 & 0.34 & 0.00 & 0.20 & 0.34 & 0.21 & 0.00 & 0.46 & 0.00 & 0.00 & 0.00 \\
14 & 0.19 & 0.38 & 0.24 & 0.24 & 0.00 & 0.00 & 0.26 & 0.00 & 0.35 & 0.00 \\
15 & 0.38 & 0.31 & 0.00 & 0.00 & 0.00 & 0.00 & 0.00 & 0.01 & 0.44 & 0.00 \\ \hline
\end{tabular}
\caption{The values of $\lambda_i$ and $p_{i,j}$ used in the example.}
\label{priors}
\end{table}

\subsection{Emulation}
\label{sec:exem}

We wish to sequence the tasks in order to minimise the expected cost and time. We do so by maximising the prior expected utility as outlined above. We have a single binary utility function for cost and time. The conditional utilities used are those in (\ref{eq:util}). Suppose that initially the decision maker felt that the utility of each was equally important. Then $q_1=1/2,q_2=1/2,q_3=0$.

For this illustrative example with $J=9$, evaluating the expected utilities  for all of the sequences takes 102 seconds in R version 3.3.1 on a machine with 16GB of RAM and an Intel Core i7-6700@3.40 GHz processor. If we increase the number of tasks this exhaustive search would take around 17 minutes for $J=10$, 3.1 hours for $J=11$, 37 hours for $J=12,$ 20 days for $J=13$, 284 days for $J=14$ and 11.7 years for $J=15.$ Thus, while we have chosen a number of tasks for which we can evaluate the success of the emulator by comparison with the exact solution in this case, for moderately larger decision problems we would be unable to solve the problem exactly.

\begin{figure}[ht]
\centering
\includegraphics[width=0.48\linewidth]{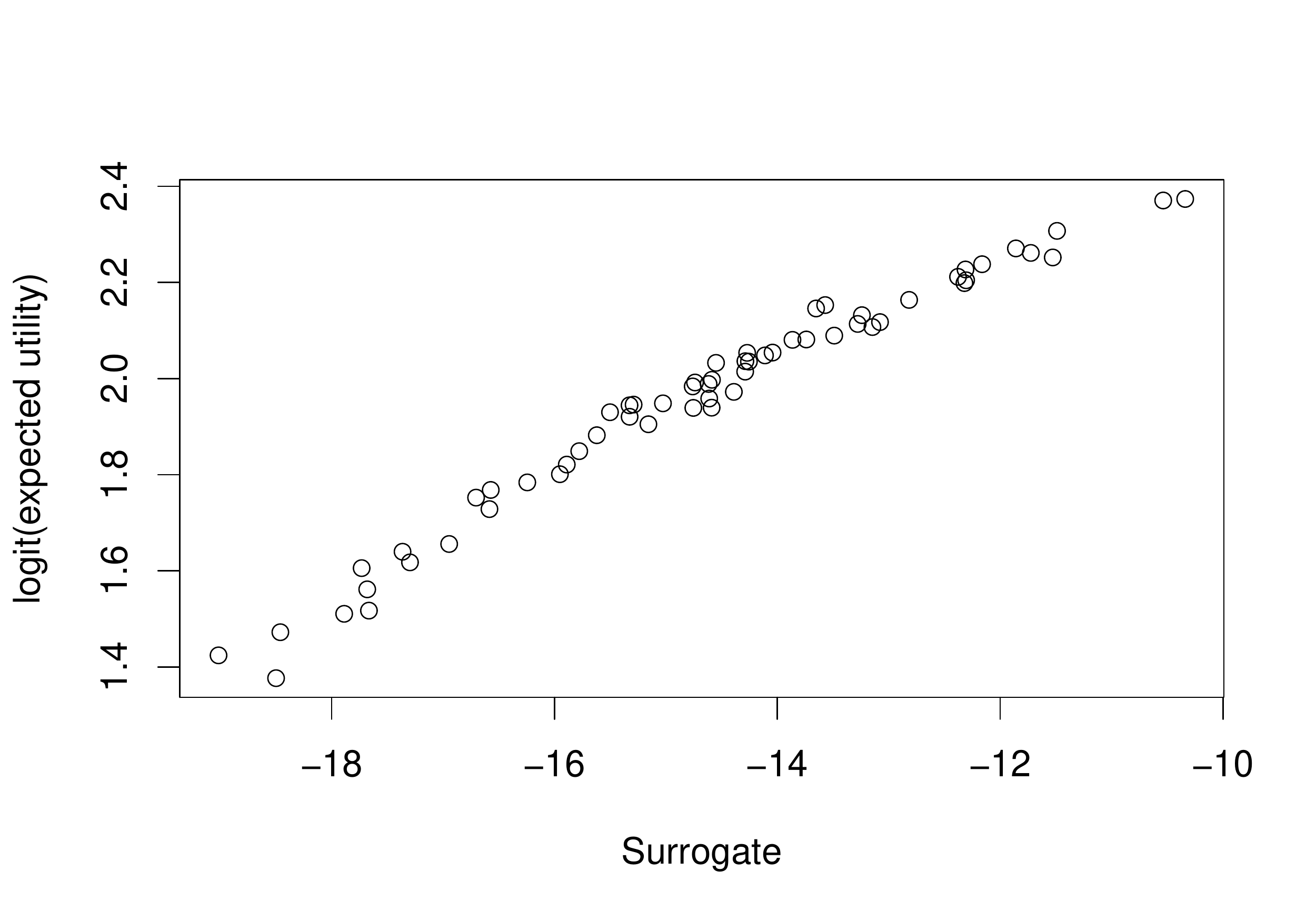}
\includegraphics[width=0.48\linewidth]{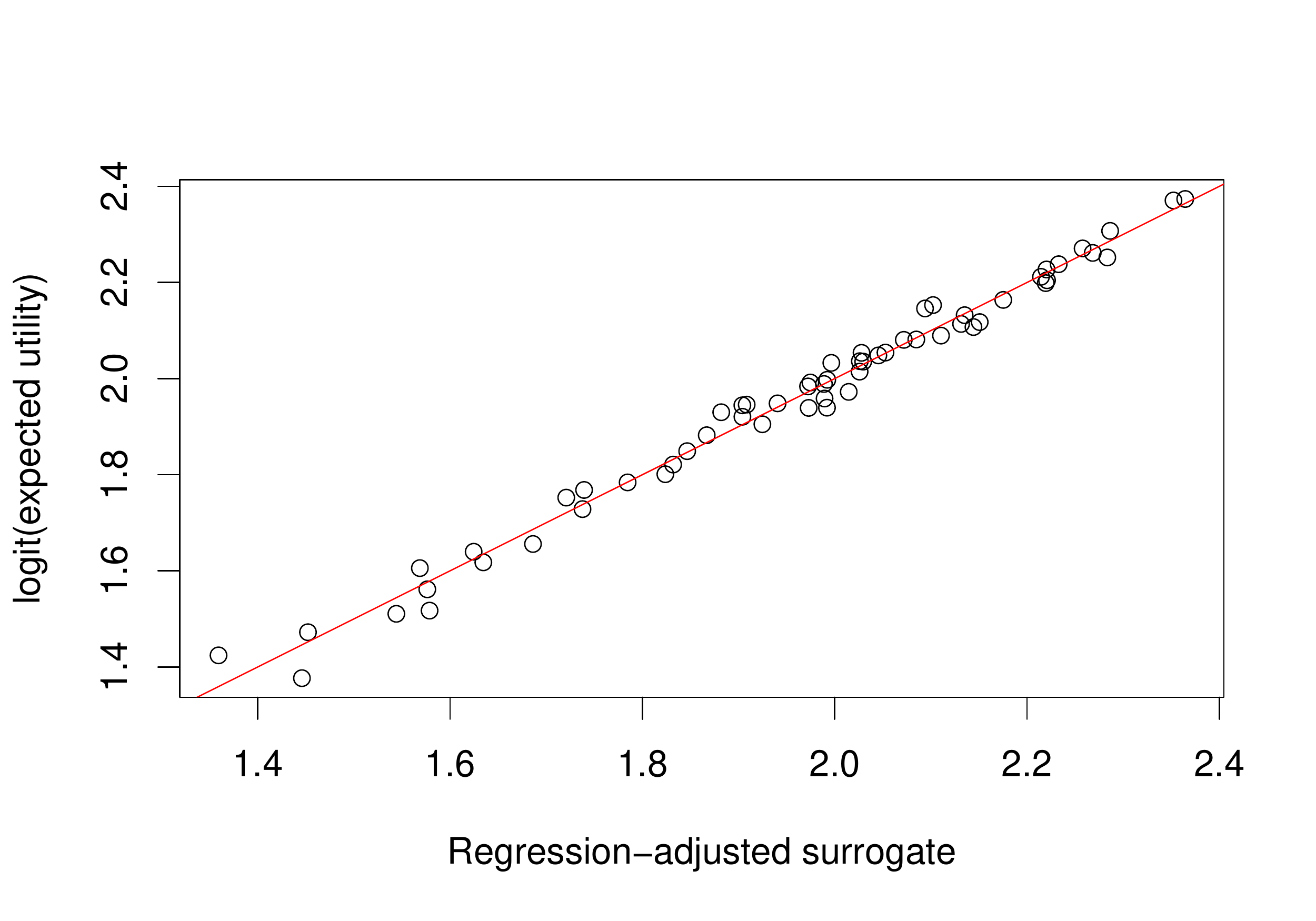}
\caption{Logit transformed expected utilities versus surrogate
  function values for the training sample of sequences under the
  initial surrogate $\hat{f}(\cdot)$ (left) and the regression-adjusted
  surrogate $f^\star(\cdot)$ (right).}
\label{Scatter}
\end{figure}

We suppose that we have a budget of $B=100$ evaluations of the
expected utility, which will be split between $N$ evaluations at the
training sample and $M$ evaluations based on the top $M$ sequences
under the emulator.  In order to focus attention on the ability of the
emulator, rather than the choice of training sample, we choose the
simplest possible design for our training sample, i.e.\ $N$ sequences
sampled uniformly from the set of all $J!$ sequences. With such a
design, the empirical results of Section~\ref{simul} suggest that the
optimal split of our budget is approximately $N=60$ and $M=40$, giving
roughly a 75\% chance of obtaining the optimal sequence. We use the
Pearson correlation as our measure of correlation $c(\vec{\psi})$,
again due to its superior empirical performance; see
Section~\ref{simul}. We first fit the initial surrogate based on the
Benter model, $\hat{f}(\cdot)$, as described in
Section~\ref{emuldesc}.  %The maximisation of the
%correlation is performed using the \verb optim function in \verb R. We
%run the optimisation from five different starting points, and choose
%the value of $\bm\psi$ which gave the maximum correlation. 
The optimal
choice for this illustrative example is
$\hat{\bm\theta}=(1.00,0.79,2.36,2.72,0.38,3.20,0.99,3.45,0.89)$ and
$\hat{\bm\alpha}=(1.37,1.43,1.58,1.31,0.57,0.18,0.03,0.13,0)$. This
gives a sample Pearson correlation coefficient of
$c(\hat{\bm\psi})=0.984$, suggesting that we may be able to use
$\hat{f}(\cdot)$ as a surrogate for the logit expected utility in a
search for the optimal sequence. It is good practice to produce a
visual check of the association between the output under the surrogate
and the logit expected utilities (to decide if the
association is likely to produce a good emulator). The values of the
logit expected utility and the surrogate function values at the 60
training points for this choice of $\bm\psi$ are given in the left hand
side of Figure \ref{Scatter} and show a good correspondence.

Whilst the estimated parameters are largely a vehicle for obtaining an
emulator, they do provide some insight into the solution of the design problem. The rank order of the $\hat{\theta}_k$ from
largest to smallest should be similar to the order of tasks with the highest value
of the surrogate function: in this case the order is $(8,6,4,3,1,7,9,2,5)$,
suggesting task 8 is performed first, followed by task 6 and so
on. The relative values of the $\hat{\theta}_k$ also provide a rough
guide to how persistent the tasks are likely to be in the sequences
with the highest values of the surrogate function; for example we can
see that $\hat{\theta}_5$ is much lower than the other parameters and
so task 5 is likely to always be scheduled last in the sequences with high
values of the surrogate function. Similarly, $\hat{\alpha}_j$
 can be interpreted as the importance of the task that is performed
 $j$th in the sequence. We see that the values for $\alpha_1$ to
 $\alpha_4$ (and possibly $\alpha_5$) are all relatively large and then the values start to tail
 off.  This suggests that the first four (or five) tasks may be the
 important tasks to schedule, perhaps due to the reliability target being
 reached by then, with the last few not as important (in which case
 the order in which the tasks are performed may have little impact on
 the expected utility). 

%\begin{comment}
%The strong positive correlation between the logit expected utility and
%$\hat{f}(\cdot)$  suggests that we may be able to use $\hat{f}(\cdot)$
%as a surrogate for the logit expected utility in a search for the
%optimal sequence.
%\end{comment}

We can evaluate the suitability of the regression-adjusted
surrogate function; see Section~\ref{sec:rasurr}. In this case we obtain
$\widehat{\vec{\beta}}=(2.31,-0.07,-0.01,0.00)$ and this gives a
sample correlation coefficient of $0.993$. The logit expected
utilities are plotted against the regression-adjusted surrogate
function values in the right hand side of Figure \ref{Scatter}. The correlation is very high for both surrogate functions, indicating
that either would be a good emulator, however, in what follows we use the
regression-adjusted surrogate as that gave the higher Pearson
correlation.  

%\begin{figure}[ht]
%\centering
%\includegraphics[height=2.5in]{Scatter2.png}
%\caption{Log surrogate function values versus logit transformed expected utilities for the training sample of sequences under the reverse Plackett-Luce model.}
%\label{Scatter2}
%\end{figure}

We evaluate the regression-adjusted surrogate function at all 9!
sequences of tasks; this takes approximately 17 seconds which is 6
times quicker than evaluating all of the expected utilities. For
increasing numbers of tasks the emulation would take around 3 minutes
for $J=10$, 31 minutes for $J=11$, 6 hours 15 minutes for $J=12$, 3.4
days for $J=13,$ 47 days for $J=14$ and 1.9 years for $J=15$. Each of
these numbers represents a significant speed up over the exhaustive
enumeration of the expected utility. However, even
  for moderate $J$ complete enumeration may be infeasible.  This
  potential drawback
  is easily addressed by the probabilistic nature of the surrogate model, as
  described in Section~\ref{sec:largeJ}.

\begin{figure}[ht]
\centering
\includegraphics[width=0.6\linewidth]{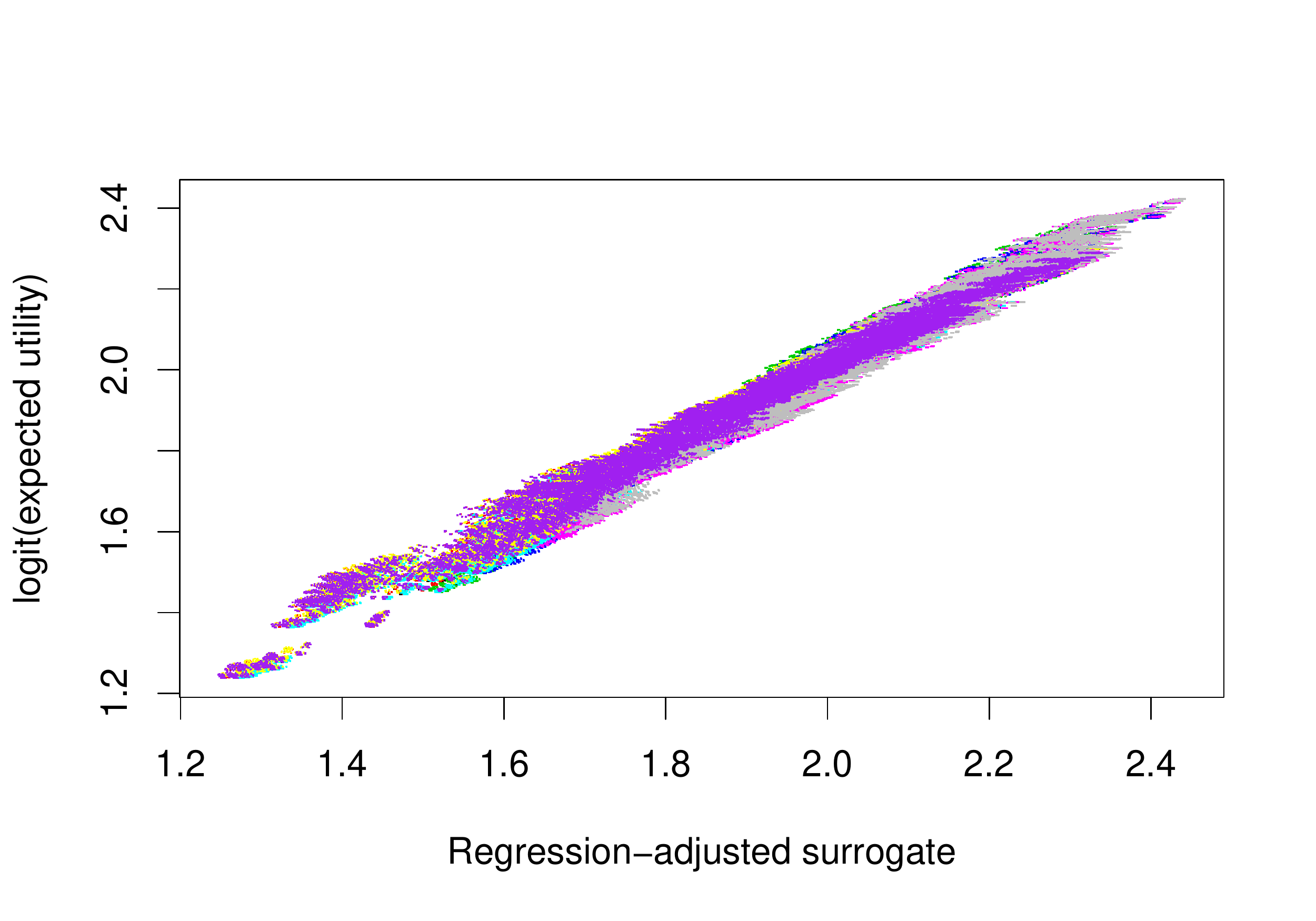}
\caption{Logit transformed expected utilities versus regression-adjusted surrogate function values for all sequences $\bm x$; the colours indicate the first task in that sequence (1=black, 2=red, 3=green, 4=blue, 5=cyan, 6=magenta, 7=yellow, 8=grey, 9=purple).}
\label{Scatter3}
\end{figure}

Figure \ref{Scatter3} plots the logit transformed expected utility
against the regression-adjusted  surrogate function for all 362,880 sequences.

The points in Figure \ref{Scatter3} are colour-coded by the first task
in each sequence. There is a clear positive relationship between the
expected utilities and the surrogate output. The Pearson correlation
coefficient is 0.990. As there is not perfect correspondence between
$U(\bm x)$ and $f(\bm x)$, we propose the remaining $B-N=M=40$ candidate sequences
which have the highest values under the emulator $f^\star(\cdot)$ and
then evaluate the expected utility at these $M$ candidate values. In
the interests of space, only the top ten sequences in terms of their values of the surrogate function are given in Table \ref{surr}.
\begin{table}[ht]
\centering\fbox{
\begin{tabular}{c|ccccccccc|c|c|c}
Sequence $(m)$ & \multicolumn{9}{|c|}{Task sequence} & $f^\star(\bm x)$ &
$\text{logit} U(\bm x)$ & $\hat{P}$ \\ \hline
 1 &   8  & 6  & 4 &  3 &  1 &  7 &  2 &  9 &  5 & 2.442344 &2.422417&0.17\\
 2 &   8  & 6  & 4 &  3 &  1 &  7 &  9 &  2 &  5 & 2.442185 & 2.422419&0.19\\
 3 &   8  & 6  & 4 &  3 &  7 &  1 &  2 &  9 &  5 & 2.442108 & 2.422063&0.21\\
 4 &   8  & 6  & 4 &  3 &  7 &  1 &  9 &  2 &  5 & 2.441948 & 2.422065&0.23\\
 5 &   8  & 6  & 4 &  3 &  1 &  9 &  2 &  7 &  5 & 2.441792 & 2.422417&0.25\\
 6 &   8  & 6  & 4 &  3 &  1 &  9 &  7 &  2 &  5 & 2.441504 & 2.422418&0.27\\
 7 &   8  & 6  & 4 &  3 &  7 &  9 &  2 &  1 &  5 & 2.441485 & 2.422040&0.29\\
 8 &   8  & 6  & 4 &  3 &  7 &  9 &  1 &  2 &  5 & 2.441180 & 2.422044&0.32\\
 9 &   6  & 8  & 4 &  3 &  1 &  7 &  2 &  9 &  5 & 2.441053 & 2.422381&0.34\\
10 &   8  & 6  & 4 &  3 &  1 &  2 &  9 &  7 &  5 & 2.440947 & 2.422380&0.36\\
\end{tabular}}
\caption{The ten sequences with the highest values of $f^\star(\bm x)$
  and their corresponding logit expected utilities. The final column
  gives an estimate $\hat{P}$ for the probability that the optimal sequence is
  to be found in the top $m$ sequences under the emulator for examples
of this type.}
\label{surr}
\end{table}
% [1] 0.17 0.19 0.21 0.23 0.25 0.27 0.29 0.32 0.34 0.36
Whilst the sequence $\bm x=(8,6,4,3,1,7,2,9,5)$ has the largest value
under the emulator it does not have the highest expected
utility. Nevertheless, these top 10 sequences all have high expected
utility. The sequence which gives the maximum expected utility out of
the $B=N+M=100$ at which the expected utility was calculated, and is
therefore our putative optimal sequence, is $\tilde{\bm
  x}=(8,6,4,3,1,7,9,2,5)$ which is ranked 2nd in terms of the
emulator. %Of course, in practice, the maximum possible expected
%utility of 1 is unlikely ever to be achieved and so we are looking
%for an expected utility of close to 1.
Also in Table~\ref{surr} we give an estimate $\hat{P}$ for the probability that the optimal sequence is
  to be found in the top $m$ sequences under the emulator for examples
of this type. These estimates are based on a regression analysis of
the simulation results of Section~\ref{simul}. 

As mentioned previously, we can evaluate all of the expected utilities for this example. The sequences ranked by the highest expected utilities are given in Table \ref{util}. In practice, these sequences would not typically be known. We show them here to investigate the ability of the surrogate function to identify the globally optimal sequence.

\begin{table}[ht]
\centering\fbox{
\begin{tabular}{c|ccccccccc|c|c}
Sequence & \multicolumn{9}{|c|}{Task sequence} & $f^\dagger(\bm x)$ & $U(\bm x)$ \\ \hline
 1 &   8 &  6 &  4 &  3 &  1 &  7 &  9 &  2 &  5 & 0.9199880 & 0.9185209\\
 2 &   8 &  6 &  4 &  3 &  1 &  7 &  9 &  5 &  2 & 0.9197039 & 0.9185209\\
 3 &   8 &  6 &  4 &  3 &  1 &  9 &  7 &  2 &  5 & 0.9199380 & 0.9185209\\
 4 &   8 &  6 &  4 &  3 &  1 &  9 &  7 &  5 &  2 & 0.9196522 & 0.9185209\\
 5 &   8 &  6 &  4 &  3 &  1 &  7 &  2 &  9 &  5 & 0.9199998 & 0.9185208\\
 6 &   8 &  6 &  4 &  3 &  1 &  9 &  2 &  7 &  5 & 0.9199592 & 0.9185208\\
 7 &   8 &  6 &  4 &  3 &  1 &  7 &  2 &  5 &  9 & 0.9196663 & 0.9185208\\
 8 &   8 &  6 &  4 &  3 &  1 &  9 &  2 &  5 &  7 & 0.9195836 & 0.9185208\\
 9 &   8 &  6 &  4 &  3 &  1 &  7 &  5 &  9 &  2 & 0.9197978 & 0.9185195\\
10 &   8 &  6 &  4 &  3 &  1 &  7 &  5 &  2 &  9 & 0.9197485 & 0.9185195\\
\end{tabular}}
\caption{The ten sequences with the highest values of $U(\bm x)$ and
  their corresponding values under $f^\dagger(\cdot)$.}
\label{util}
\end{table}

The optimal sequence of tasks, that with the largest expected
utility, is $\hat{\bm x}=(8,6,4,3,1,7,9,2,5)$. This matches
$\tilde{\bm x}$, the putative optimal sequence which was highlighted
in our candidate list.   In this
illustrative example, we have found the optimal sequence using only
$B=100$ evaluations of the expensive expected utility. Our simulation
results in Section~\ref{simul} suggest that in similar examples, with this
choice of $N$ and $M$ we would obtain the optimal sequence roughly
75\% of the time. Even when we do not find the optimal sequence in the
$B$ evaluations, we still find sequences with close to optimal values
of the expected utility. For example, in this illustrative example, the sequence with maximum value of the
surrogate function has the 5th highest expected utility and matches
the optimal sequence in the first five performed tasks. So, even if we
set $M=1$ we would end up with only a marginally sub-optimal
sequence.

%Remove this paragraph as we would never really just sample sequences
%at random like this

%The advantage of using the emulator becomes clear when we consider the
%expected utilities of the sequences from the training sample of 60
%randomly chosen sequences. Of these, the estimate of the optimal
%sequence would have been $\bm x=(8,6,3,1,4,9,2,5,7)$, which has an
%expected utility of 0.916, much smaller than any of those in the top $M$ based on the emulator. In fact there are 1684 sequences which have a higher expected utility than this sequence.

\begin{figure}[ht]
\centering
\includegraphics[width=0.48\linewidth]{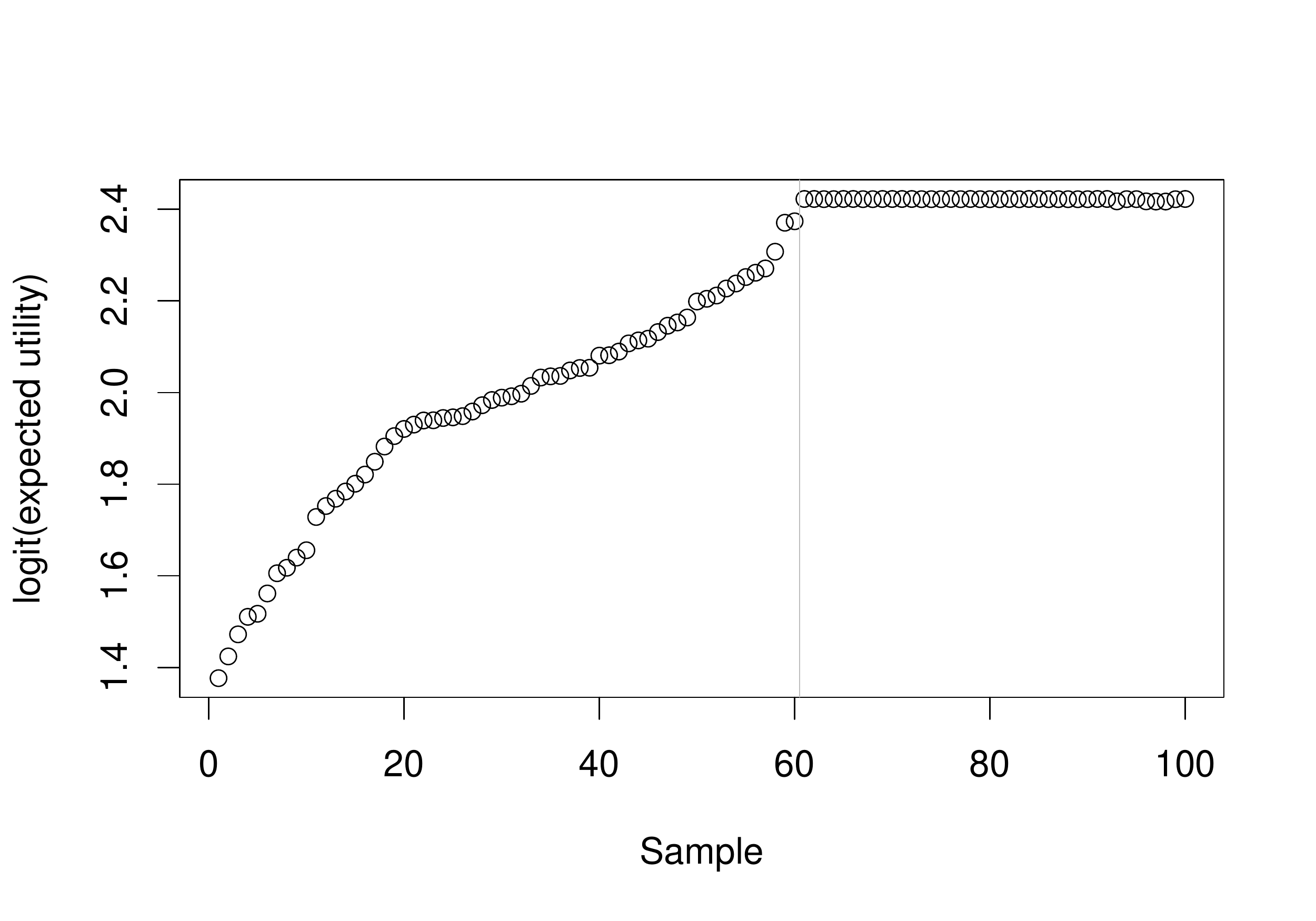}
\includegraphics[width=0.48\linewidth]{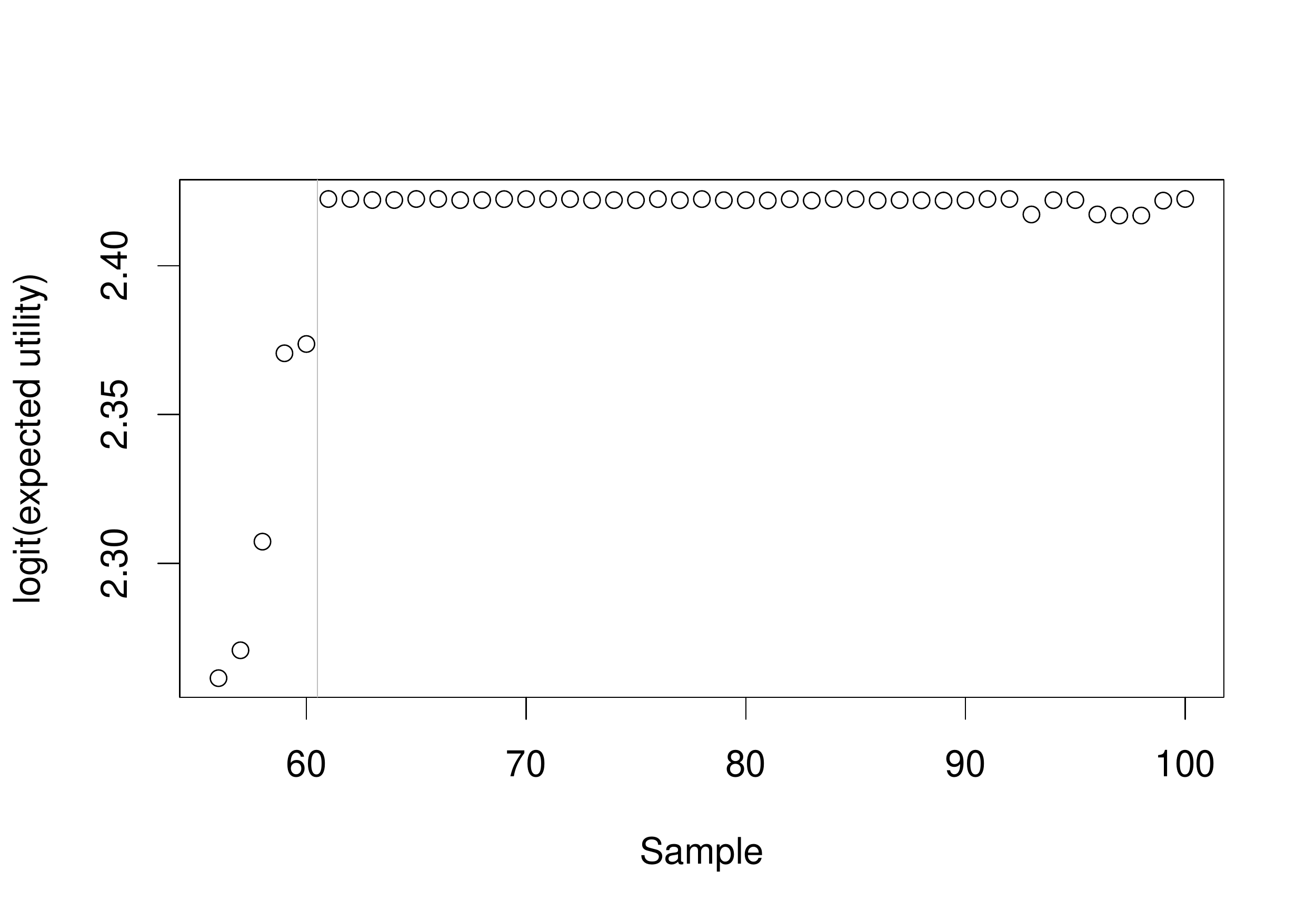}
\caption{Logit transformed expected utilities versus sample number. In
  each plot the points to the left of the vertical line represent the
  training samples (ordered by increasing expected utility) and the
  points to the right of the line correspond to the 
  $M$ sequences suggested by the emulator in decreasing order of the regression-adjusted
  surrogate function value. The left-hand plot shows all $B=100$ samples. The
  right-hand plot focuses on the top 5 from the training sample and
  the $M$ sequences proposed from the emulator.}
\label{fig:diagnostic}
\end{figure}

As illustrated above, in this example we have been  
  able to evaluate  the expected utility at all possible sequences
  of tasks and therefore been able to identify the optimal sequence,  but our
  methodology is designed for scenarios where this is not possible. In
  such situations, a practitioner will not know whether they have
  obtained the optimal sequence, or whether their putative optimal
  sequence is close to the optimal. We therefore recommend a simple graphical
  diagnostic plot along the lines of that in
  Figure~\ref{fig:diagnostic}. 
This compares graphically the logit transformed expected utilities in the training sample of
size $N$ with
those in the evaluation sample of size $M$. From such a plot, a
practitioner can see the extent to which the $M$ emulator-proposed sequences all
have much larger expected utilities than those in the training sample
and also the extent to which the expected utilities of the  $M$
emulator-proposed sequences are all very similar. This is reassuring
as it suggests that the putative optimal sequence is probably close to
optimal (in terms of the overall variation in logit expected utilities
that are observed in the training sample). So, whilst, in a real
example, there are no
guarantees that the optimal sequence has been found, such a graphical
diagnostic can provide some reassurance that the putative optimal
sequence is
satisfactory.

%Add in about stability after added to supplementary material.
%A suggestion of one of the referees of splitting the training sample
%and fitting separate emulators to the sub-samples in order to check
%the stability of the emulators and hence the putative optimal sequences
%has been investigated but was found not to provide a significant
%improvement to the procedures outlined above; see the Supplementary
%Material for details.  

%prefer this -can easily be deleted
As suggested by one of the referees, an alternative strategy for checking the stability of the
emulator and of the putative optimal sequence is to split the
training sample into sub-samples and perform the emulator
fitting/exploration process on each sub-sample. Our experience, as
reported in the Supplementary Material, is that not much insight is gained by
this splitting of the training sample and so we have chosen not to pursue that
procedure here.

\subsection{Results}

The optimal ordering of tasks has an expected utility of 0.919 and corresponds to the sequence $(8,6,4,3,1,7,9,2,5)$. That is, we first carry out task 8 and then if we haven't met the reliability target we move on to task 6. If we still haven't met the reliability target we then carry out task 4, etc. 

We consider the costs, times and probabilities of reaching the target reliability following each task. They are given in Table \ref{tab:seq}.

\begin{table}[ht]
\centering
\fbox{
\begin{tabular}{c|ccccccccc}
Task & 8 & 6 & 4 & 3 & 1 & 7 & 9 & 2 & 5 \\ \hline
Cost & 7 &  23 &  31 &  37 &  48 &  60 &  66 & 115 & 132 \\
Time & 10 & 24 & 26 & 28 & 29 & 48 & 61 & 63 & 73 \\
Probability & 0.00 & 0.08 & 0.48 & 0.87 & 0.98 & 0.99 & 1.00 & 1.00 & 1.00 \\
\end{tabular}}
\caption{Various quantities of interest broken down by task in the optimal sequence.}
\label{tab:seq}
\end{table} 

%Plots of the probability of reaching the target after each task by the number of tasks completed (left) and the time and financial costs against the probability of reaching the target after each task (right) are given in Figure \ref{final}. In the right-hand plot financial costs are in red and time on test is in blue.

%\begin{figure}[ht]
%\centering
%\includegraphics[height=2.5in]{Final.png}
%\caption{The probability of reaching the target reliability for a given number of tasks (left) and the proportion of total financial cost (red) and time on test (blue) against the probability of reaching the target reliability (right).}
%\label{final}
%\end{figure}

From the table we see that it is not likely that we will have to
perform all of the tasks to achieve the target reliability. It is
likely that only 4 or 5 tasks will be needed and that we can reach the
reliability target with a spend of 37-48 and a time of 28-29. Note
that this coincides with our interpretation of the $\alpha$ 
parameters; the first 4 or 5 tasks have relatively large values
of $\alpha$ compared to the other tasks. We see that we are likely to spend a larger proportion of our total time than our total budget for testing.

\section{Simulation studies}
\label{simul}

\subsection{Choice of surrogate model and correlation function}
\label{sec:emcomp}

A simulation study was performed to assess the performance of several
surrogate models based on parametric probabilistic models for
permutations: the Benter model (B) as described in
Section~\ref{emuldesc}; the Plackett-Luce model (PL) which is obtained
from the Benter model by setting $\vec{\alpha}=\vec{1}$; and the
reverse Plackett-Luce model (RPL) which is obtained via the PL model
evaluated at the reverse sequence of tasks. We did not include the
regression-adjustment in the simulation study since it can be
applied to any surrogate function and will improve upon it. A range of correlation functions
(Pearson, Spearman and Kendall) were considered as the objective
function for the fitting procedure for each of the putative surrogate
models.  We have considered numbers of tasks $J$ ranging from 6 to 10
and three sets of trade-off parameters: (a) $q_1=1/2, q_2=1/2, q_3=0$,
(b) $q_1=1/3, q_2=1/3, q_3=1/3$ and (c) $q_1=2/3, q_2=2/3, q_3=-1/3$.
The problem set-up follows that in Section~\ref{example} in which we
set $\epsilon_i=\epsilon=0.02$, $t=100$, $R_0=0.8$, $T_0=90$. To focus
attention on the performance of the emulator, rather than the
effect that the design of the training samples has on performance, we use the randomly sampled design strategy that
was used for the illustrative example in Section~\ref{example}. In
this way we separate the potential capability of each emulator from
the issue of training sample design.

For each combination of number of tasks $J$, trade-off parameters $\vec{q}$ and initial random sample
size $N$, 100 sets of expected utilities for all $J!$ sequences were
generated.  For each of these 100 simulations we fitted each of the
nine combinations of surrogate model and correlation function based on the
initial sample of $N$ sequences. We evaluated the surrogate
function at all $J!$ possible sequences.  %Through the simulation
%study we aim to provide guidance on choice of sample sizes $N$ and
%$M$ for this simple choice of training sample, as well as providing
%guidance of the relative merits of the various surrogates.

%The simulations were
%performed in \textsf{R} version 3.2.3 on a modern desktop computer. 
We focus on the results for $J=9$ tasks and trade-off parameters (b);
the differences for the other combinations of trade-off parameters were
minimal and other values of $J$ from 6 to 10 gave broadly similar
results, though the differences between the performance of the
surrogate models and correlation functions increase as $J$ increases.

Table~\ref{tab:probtopJ9} shows the estimated probability that the
optimal sequence in terms of expected utility is contained in either the $N$
initial sequences or the further $M$ sequences based on the best $M$
sequences in terms of the surrogate function, for a range of values of
$N$ and $M$. 
%\vspace{-0.5in}

\singlespacing

\begin{table}[h!]
\centering
{\scriptsize
\begin{tabular}{cccccccc}
\hline
\hline
       &         &            &\multicolumn{5}{c}{$M$} \\
\cline{4-8} 
 $N$   &  Correlation & Model & 10 & 20 & 50 & 100 & 200 \\
\hline
$25$   &  Pearson & PL & 0.05 & 0.06 & 0.10 & 0.19 & 0.27\\
       &       & RPL & 0.09 & 0.15 & 0.25 & 0.33 & 0.50\\
       &       & B & 0.19 & 0.30 & 0.46 & 0.60 & 0.73\\
%       &       & RAB &  &  &  &  & \\
\cline{2-8} 
       & Spearman & PL & 0.04 & 0.05 & 0.05 & 0.05 & 0.06\\
       &       & RPL & 0.01 & 0.02 & 0.06 & 0.10 & 0.14\\
       &       & B & 0.00 & 0.00 & 0.01 & 0.01 & 0.01\\
%       &       & RAB &  &  &  &  & \\
\cline{2-8} 
       & Kendall & PL & 0.01 & 0.01 & 0.02 & 0.03 & 0.06\\
       &       & RPL & 0.01 & 0.03 & 0.04 & 0.07 & 0.11\\
       &       & B & 0.00 & 0.00 & 0.00 & 0.00 & 0.01\\
%       &       & RAB &  &  &  &  & \\
\hline
$50$   &  Pearson & PL & 0.11 & 0.19 & 0.35 & 0.54 & 0.70\\
       &          & RPL & 0.26 & 0.38 & 0.56 & 0.68 & 0.80\\
       &          & B & 0.40 & 0.51 & 0.73 & 0.85 & 0.95\\
%       &          & RAB &  &  &  &  & \\
\cline{2-8} 
       & Spearman & PL & 0.02 & 0.04 & 0.06 & 0.17 & 0.27\\
       &          & RPL & 0.05 & 0.09 & 0.17 & 0.22 & 0.29\\
       &          & B & 0.03 & 0.07 & 0.13 & 0.23 & 0.26\\
%       &          & RAB &  &  &  &  & \\
\cline{2-8} 
       & Kendall & PL & 0.05 & 0.08 & 0.12 & 0.16 & 0.28\\
       &         & RPL & 0.06 & 0.18 & 0.22 & 0.28 & 0.35\\
       &         & B & 0.06 & 0.10 & 0.13 & 0.19 & 0.25\\
%       &         & RAB &  &  &  &  & \\
\hline
$75$   &  Pearson & PL & 0.15 & 0.29 & 0.49 & 0.67 & 0.77\\
       &          & RPL & 0.35 & 0.46 & 0.68 & 0.77 & 0.89\\
       &          & B & 0.49 & 0.66 & 0.87 & 0.96 & 0.99\\
%       &          & RAB &  &  &  &  & \\
\cline{2-8} 
       & Spearman & PL & 0.06 & 0.11 & 0.17 & 0.26 & 0.35\\
       &          & RPL & 0.16 & 0.23 & 0.29 & 0.43 & 0.56\\
       &          & B & 0.17 & 0.21 & 0.32 & 0.42 & 0.54\\
%       &          & RAB &  &  &  &  & \\
\cline{2-8} 
       & Kendall & PL & 0.04 & 0.09 & 0.19 & 0.26 & 0.34\\
       &         & RPL & 0.10 & 0.19 & 0.31 & 0.42 & 0.50\\
       &         & B & 0.10 & 0.20 & 0.32 & 0.38 & 0.47\\
%       &         & RAB &  &  &  &  & \\
\hline
$100$  &  Pearson & PL & 0.15 & 0.34 & 0.48 & 0.67 & 0.83\\
       &          & RPL & 0.37 & 0.59 & 0.81 & 0.92 & 0.98\\
       &          & B & 0.54 & 0.72 & 0.94 & 0.98 & 1.00\\
%       &          & RAB &  &  &  &  & \\
\cline{2-8} 
       & Spearman & PL & 0.11 & 0.15 & 0.24 & 0.34 & 0.41\\
       &          & RPL & 0.20 & 0.28 & 0.46 & 0.59 & 0.69\\
       &          & B & 0.21 & 0.33 & 0.50 & 0.61 & 0.68\\
%       &          & RAB &  &  &  &  & \\
\cline{2-8} 
       & Kendall & PL & 0.07 & 0.13 & 0.24 & 0.31 & 0.39\\
       &         & RPL & 0.14 & 0.21 & 0.34 & 0.43 & 0.60\\
       &         & B & 0.20 & 0.28 & 0.43 & 0.55 & 0.68\\
%       &         & RAB &  &  &  &  & \\
\hline
$200$  &  Pearson & PL & 0.36 & 0.57 & 0.72 & 0.87 & 0.96\\
       &          & RPL & 0.62 & 0.86 & 0.95 & 0.98 & 1.00\\
       &          & B & 0.69 & 0.85 & 0.91 & 0.97 & 1.00\\
%       &          & RAB &  &  &  &  & \\
\cline{2-8} 
       & Spearman & PL & 0.22 & 0.31 & 0.42 & 0.55 & 0.70\\
       &          & RPL & 0.26 & 0.45 & 0.67 & 0.77 & 0.87\\
       &          & B & 0.56 & 0.72 & 0.88 & 0.96 & 0.99\\
%       &          & RAB &  &  &  &  & \\
\cline{2-8} 
       & Kendall & PL & 0.15 & 0.25 & 0.39 & 0.53 & 0.70\\
       &         & RPL & 0.26 & 0.38 & 0.56 & 0.67 & 0.81\\
       &         & B & 0.53 & 0.65 & 0.79 & 0.86 & 0.90\\
%       &         & RAB &  &  &  &  & \\
\hline
\hline
\end{tabular}
}
\caption{Estimated probability from 100 simulations that the optimal
  sequence is found in an initial random sample of $N$ sequences
  followed by the top $M$ sequences under various surrogates:
  PL=Plackett-Luce, RPL=Reverse Plackett-Luce, B=Benter.} %PL-P and
  %RPL-P denote
  %the Plackett-Luce and reverse Plackett-Luce surrogates fitted using
  %Pearson correlation, RPL-K denotes the reverse Plackett-Luce
  %surrogate fitted using Kendall correlation. The three choices of
  %trade-off parameters are (1) $p_1=1/2, p_2=1/2, p_3=0$, (2) $p_1=1/3,
  %p_2=1/3, p_3=1/3$ and (3) $p_1=2/3,
  %p_2=2/3, p_3=-1/3$.}
\label{tab:probtopJ9}
\end{table}

\doublespacing

The results in Table~\ref{tab:probtopJ9} show that using
Pearson correlation for the objective function is uniformly better
than using the other two correlation functions. Within the results for
the Pearson correlation, the surrogate based on the Benter model
($\hat{f}(\cdot)$), is almost
uniformly better than the other two models (PL and RPL), and provides
a noticeable improvement for small $N$ and $M$. The results justify our
preference for the Benter model over the PL or RPL models as the basis
for a parametric surrogate. Due to its dominance in the simulations, in
what follows we restrict attention to the combination of Benter and Pearson.

\subsection{Guidelines on choice of $\boldsymbol{N}$ for a given budget
  $\boldsymbol{B}$}
\label{sec:guidelinesN}

Table~\ref{tab:probtopJ9}
provides some guidelines on the choices of $N$ and $M$ for problems
such as this, which are 
useful if we have a specific budget of $B=N+M$ sequences for which we
can calculate the expected utility. For example, with a budget of $B=150$
sequences, the probability that we obtain the optimal sequence is
approximately 85\% when $N=50, M=100$, whereas
with $N=100, M=50$ this probability is approximately 94\%. 

The results in Table~6 relate to one specific
  ``problem scenario'' as defined through the values of $J, I, R_0,
  \epsilon, \lambda, p, q $ and so on, and are perhaps too narrowly
  focused to provide generic advice.  A more extensive simulation
  study was carried out in order to offer more generic guidance on the
  choice of $N$ for a given $B$, over a broad range of ``problem
  scenarios''.  A total of 10000 randomly generated ``problem
  scenarios'' were analysed in the second simulation study and the
  results are described in detail in the Supplementary Material.
  Specifically, we fit a logistic regression model with response
  variable an indicator of whether the optimal sequence was found for
  that problem scenario  and with explanatory variables equal to $N$, $B$ and
  functions thereof. This approach allows us to predict the
  probability of finding the optimal sequence for combinations of $N$
  and $B$, to approximate the optimal choice of $N$ for a given budget
  $B$ and predict the probability of obtaining the optimal sequence
  for that choice. Based on the results of this second simulation
  study we would recommend the \textit{rule of thumb} of setting
  $N=M$, or, in other words, setting $N=B/2$ (to the nearest integer).
  This provides a simple, default first choice in the absence of other
  information.  Our simulations suggest that this is only marginally
  suboptimal, as well as being very simple and easy to remember.

Additionally, we have used the
simulation results from Table~\ref{tab:probtopJ9} to determine an
optimal choice of $N$ for a given budget $B$ for the specific
``problem scenario'' considered therein.  Specifically, following the
logistic regression procedure outlined in the Supplementary Material, the results suggest that, if $B=100$, which was used in the illustrative example of Section~\ref{example},
the optimal choice of $N$ is 58 (and therefore $M=42$); this
gives a probability of obtaining the optimal sequence of about
$75\%$. This is very similar to our generic \textit{rule of thumb}
which  would suggest a 50-50 split between $N$ and $M$  and a
probability of obtaining the optimal sequence of just over $70\%$ (as
determined by the results in the Supplememtary Material).

\begin{comment}
when $B=100$ we should choose $N=58$ and this would give 
for a given budget $B$ choosing $N \simeq 8+0.5\times B$ is optimal in
problem scenarios of this type.
With the optimal choice of $N$, the estimated probability of observing
the optimal sequence in the $B=N+M$ sequences is given in
Figure~\ref{fig:probbudgetsims1J9}.
\begin{figure}[h!]
\centering
\includegraphics[width=0.6\linewidth]{../../graphics/poptinB.pdf}
\caption{Estimated probability that the optimal sequence is found
  against budget, $B$ for $N$ and $M$ chosen optimally for the
  emulator based on the Benter model with maximised Pearson correlation.}
\label{fig:probbudgetsims1J9}
\end{figure}
For example, if $B=100$, which was used in the illustrative example of Section~\ref{example},
the optimal choice of $N$ is 58 (and therefore $M=42$); this
gives a probability of obtaining the optimal sequence of about $75\%$.
\end{comment} 

\subsection{Median rank performance}
\label{sec:suboptimal}

Whilst we may not obtain the optimal sequence in our $N+M$ sequences
it is interesting to know how the best ranked sequence in our sample
compares to the full set of sequences. In each of our simulations in
the first simulation study (detailed in Section~\ref{sec:emcomp}) we
computed the rank in terms of expected utilities of the highest ranked sequence
out of our sample of $N+M$ sequences for the values of $N$ and
$M$. The medians over 100 simulations of these ranks are given in
Table~\ref{tab:medianrankJ9}.
\begin{table}[h!]
\centering
{\scriptsize
\begin{tabular}{rccccc}
\hline
\hline
           &\multicolumn{5}{c}{$M$} \\
\cline{2-6} 
 $N$    & 10 & 20 & 50 & 100 & 200 \\
\hline
$25$    & 93  & 37 & 13   & 1    &  1   \\ 
$50$    & 37  &  1 &  1   & 1    &  1   \\ 
$75$    &  3  &  1 &  1   & 1    &  1   \\ 
$100$   &  1  &  1 &  1   & 1    &  1   \\ 
$200$   &  1  &  1 &  1   & 1    &  1   \\ 
\hline
\hline
\end{tabular}
}
\caption{Median from 100 simulations of the rank in terms of expected utilities
of the highest ranked sequence out of our sample of $B=N+M$ sequences
for the emulator based on the Benter model with maximised Pearson correlation.} %PL-P and
  %RPL-P denote
  %the Plackett-Luce and reverse Plackett-Luce surrogates fitted using
  %Pearson correlation, RPL-K denotes the reverse Plackett-Luce
  %surrogate fitted using Kendall correlation. The three choices of
  %trade-off parameters are (1) $p_1=1/2, p_2=1/2, p_3=0$, (2) $p_1=1/3,
  %p_2=1/3, p_3=1/3$ and (3) $p_1=2/3,
  %p_2=2/3, p_3=-1/3$.}
\label{tab:medianrankJ9}
\end{table}
We see that even in very small sample sizes we do well at choosing a
good sequence, and for most combinations of $N$ and $M$ we get the
optimal sequence at least 50\% of the time.

%Table~\ref{tab:medianrankJ9} shows a similar picture to that of
%Table~\ref{tab:probtopJ9}. In general there is little variability
%between the results for different sets of trade-off parameters. There are potentially
%big differences between choice of method, with RPL-K consistently the
%worst method. Generally RPL-P outperforms PL-P, especially for larger
%$N$ and $M$. For some small values of $N$ and $M$, PL-P is marginally
%better than RPL-P.

%We do not show a similar table for the actual expected utilities; it may be
%that even the RPL-K method gives acceptable expected utilities but we would
%always seek the method that would give us the best performance in
%terms of rankings, provided that the computational cost was the same,
%which it is here. 

%As noted above, we see a similar pattern for other values of $J$:
%$J=6,7,8$. The differences are less marked since there are fewer
%permutations so these values of $N$ cover a bigger proportion of the
%possible permutations. There is some evidence that PL-P is better than
%RPL-P for $J=8$. DO WE NEED TO MENTION THIS HERE - SORT OF COVERED
%EARLIER - POSSIBLY DELETE THIS PARAGRAPH?

\subsection{Dealing with  large numbers of permutations}
\label{sec:largeJ}

In our examples and simulation studies we have considered a fairly small
number of tasks (up to $J=10$) so that we could evaluate the expected
utility for all $J!$ sequences and solve the decision problem
exactly. This also means that we were able to evaluate the surrogate
function at all possible sequences and find the top $M$ candidates for
evaluating the expected utility. If we were interested
in sequencing a larger number of tasks, say $J=20$, it may be
infeasible to evaluate the surrogate function at all $J!$ sequences.
One advantage of an emulator based on a probabilistic model is
that we can sample a large number of sequences (though considerably
less that $J!$) from the fitted Benter model, evaluate the surrogate
function at these values, and then pick the top $M$ as the candidates
to evaluate the expected utility. This simulation-based approach
should scale up well to very large $J$, although such values are
unlikely in reliability growth.

\section{Summary and further work}
\label{conc}

We have considered the emulation of utility functions for
permutations. We proposed an approach based on the Benter model and
have shown this can provide a good approximation in reliability growth
decision making. A simple extension would be to consider the case
where some tasks can be carried out in parallel.  

Our proposed emulator is a unimodal function of its inputs. In the
illustrative example of Section~\ref{example} the utility function was
unimodal, and for the types of reliability growth tasks we consider
here we expect the utility function to be largely unimodal, and so our
proposed emulator is ideally suited. More generally, for
mutli-modal utility functions, our proposed emulator will emulate
around only one of the modes. As such, the development of a
multi-modal surrogate based on a mixture of Benter models
\citep{GormleyM08jasa,MollicaT14} may be worth investigating.

%There is potential to consider Gaussian process emulation. This would
%have advantages in the evaluation of the uncertainty on the
%approximation [OUR EMULATOR CAN ALSO DO THIS - BUT PERHAPS NOT AS WELL
%IN THAT IT DOESN'T INTERPOLATE - GPs MIGHT ALSO HANDLE MULTIMODALITY
%SO THAT COULD BE AN ADVANTAGE]. The crucial step would be to choose
%an appropriate distance measure between permutations. Initial
%investigations indicate that the Kendall distance would not be
%suitable for reliability growth decision making. 
We
chose the training set of $N$ sequences randomly. Future work would be
to choose this training sample informatively, to focus the training
sample in the areas of high expected utility. The ratio of benefit to
cost of tasks would be useful here. The approach taken for reliability growth, maximising the expected
utility of a multi-attribute utility function, can have wider use in
problems in which actions are to be ordered. A future direction is to
use the approach in optimal sequencing of actions in project risk
management.

We emphasise that the results presented in
Sections~\ref{example} and \ref{simul} are based on a simple random sample
from the set of all sequences and that other more intelligent ways to explore
the design space may lead to different results in terms of optimal
choices of $N$ and $M$. Nevertheless this
procedure allows us to assess the performance of the different
surrogates and we suspect
that the main messages from our simulation study --- that the Benter
model is preferred to the PL or RPL models, and the Pearson
correlation is the superior form of objective function --- will
persist with other types of training sample.  This has been our
experience in some preliminary investigations. Our proposed procedure has similarities to an efficient
single-step cross-entropy optimization
algorithm~\citep{RubinsteinK04}. Simulation results (given in the Supplementary Material) using a training sample derived from two steps of the
cross-entropy optimization algorithm (that is, with training sequences
sampled from the fitted model from the first stage) provide similar
conclusions to those presented here. The results in terms of the 
probability of discovering the optimal sequence in the $N+M$ sequences
are no better than those presented here for large values of $N$ and
$M$, and are typically poorer for small values of $N$ and $M$. A more
detailed investigation of design issues is the subject of future
work.

%Say something about searching sequence space more efficiently (intelligently).

\section*{Supplementary Materials}

\begin{description}

\item[Additional Details:] This file gives additional details of the work in the paper, specifically: proofs of Propositions 1 and 2, a derivation of the prior expectation of the reliability under planned
development tasks, an example of the theoretical results, an illustrative example of splitting the training sample, a simulation study giving guidance on the choice of $N$ and $M$ and simulations of a cross entropy optimization-based training sample: (PDF)

\end{description}

\section*{Acknowledgements}

We would like to thank two reviewers, the editor and an associate editor for comments and suggestions which improved the paper.

{\footnotesize
\setstretch{1}
\setlength{\bibsep}{0.0pt}
\bibliographystyle{plainnat}
\bibliography{sequencing}}

\end{document}